\documentclass{article}
\usepackage[utf8]{inputenc}

\title{A Marauder’s Map of\\ Security and Privacy in Machine Learning:\\ \large An overview of current and future research directions for making machine learning secure and private\footnote{This report summarizes the keynote presented by the author in October 2018 at AISec (colocated with ACM CCS) on security and privacy in machine learning.}.}
\author{Nicolas Papernot\\Google Brain\\papernot@google.com}
\date{}

\usepackage{comment}
\usepackage{graphicx}

\begin{document}

\maketitle

\begin{abstract}
There is growing recognition that machine learning (ML) exposes new security and privacy vulnerabilities in software systems, yet the technical community's understanding of the nature and extent of these vulnerabilities remains limited but expanding. In this talk, we explore the threat model space of ML algorithms through the lens of Saltzer and Schroeder's principles for the design of secure computer systems. This characterization of the threat space prompts an investigation of current and future research directions. We structure our discussion around three of these directions, which we believe are likely to lead to significant progress. The first encompasses a spectrum of approaches to verification and admission control, which is a prerequisite to enable fail-safe defaults in machine learning systems. The second seeks to design mechanisms for assembling reliable records of compromise that would help understand the degree to which vulnerabilities are exploited by adversaries, as well as favor psychological acceptability of machine learning applications.  The third pursues formal frameworks for security and privacy in machine learning, which we argue should strive to align machine learning goals such as generalization with security and privacy desiderata like robustness or privacy. Key insights resulting from these three directions pursued both in the ML and security communities are identified and the effectiveness of approaches are related to structural elements of ML algorithms and the data used to train them. We conclude by systematizing best practices in our community. 
\end{abstract}

\section{Introduction}

Security and privacy research in machine learning seeks to identify the degree to which learning-based systems are exposed to adversarial manipulation as well as how their behavior can be more robust to such manipulations. In this talk, we consider the security and privacy of machine learning (ML) systems rather than the application of ML to solve existing security or privacy problems. In essence, we derive insights from asking the question of how ML security and privacy relates to traditional computer security and privacy.\footnote{In the following, security is used to refer to both security and privacy.} 

Just like traditional computer security, ML security shares similarities with real-world security. As pointed out by Butler Lampson~\cite{lampson2004computer}, no real-world system is perfectly secure (e.g., it's easy to break in someone's house or forge their credit card) and therefore real-world security is all about raising the threshold for an attack to be successful so as to balance the cost of protection with the cost of recovering from an attack. Computer security is no different.

In this presentation, we ask whether machine learning offers a novel paradigm for building systems that reduces the need for this periodical reevaluation of protection and recovery costs. In other words, is machine learning also necessarily subject to the arms race found in many areas of computer security?  They are indeed some important differences between ML models and traditional computer systems. For instance, ensembling different learning systems is one of the ways to improve ML security (e.g., having an ensemble of learners vote for a prediction can make this prediction private) while combining computer systems with different architectures or configurations introduces complexity that is generally harmful to traditional computer security (e.g., running different operating system versions in a computer network). 

A conjecture on the existence of systematic approaches to security and privacy in ML systems concludes this presentation. In particular, we emphasize that such approaches, should they exist, must take care to align security and privacy goals with ML goals such as generalization.

\paragraph{Note.} This presentation is not intended to serve as a comprehensive
review of the field. In particular, we do not follow the classic presentation of this field that opposes training time adversaries to those that operate at test time, while taxonomizing attack goals under confidentiality, integrity or availability. Interested readers are referred to existing surveys covering ML security and privacy comprehensively through the lens of this classic threat model~\cite{papernot2018sok, biggio2018wild}.

\section{Why study ML security and privacy?}

Just like programs are built upon basic programming abstractions, almost all machine learning algorithms were designed with the  assumption that the model's training and test distributions are identical: points that the model will be deployed on are sampled from the same distribution than the distribution of points it was trained on. This assumption is valid for many successful applications of ML (e.g., machine translation or playing the game of Go). 

\begin{figure}[t]
\centering
\includegraphics[width=0.75\textwidth]{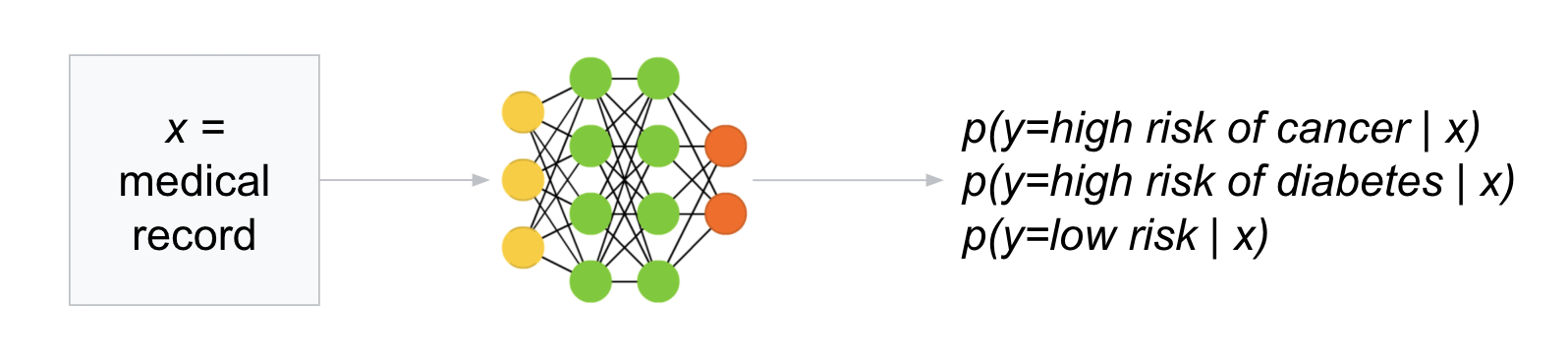}
\caption{Example of a ML-based risk model for a medical insurance provider.}
\label{fig:risk-model}
\end{figure}

To understand how this assumption influenced the design of machine learning algorithms, take a look at how current neural network classifiers are typically trained: i.e., a softmax combined with the cross-entropy loss. This loss makes the implicit assumption that all training points are completely unambiguous and should be classified with 100\% confidence in the class that is indicated by their label. For instance, Figure~\ref{fig:risk-model} depicts an example model that could be used by a medical insurance provider to estimate the risk of issuing an insurance policy to an individual. The model is trained making the assumption that all individual will fall under one of the three categories: high risk of cancer, high risk of diabetes or low risk. 
The softmax also contains exactly as many possible outputs as were included in the training data. Hence, if the model is presented with an input that does not belong to any of the classes it was trained on, it will nevertheless be forced to make a prediction in at least one of the classes it was trained on. Indeed, no process is put in place to verify inputs before
they are presented to a ML model. For instance, the risk model mentioned above would be forced to make an erroneous prediction if presented with the medical records of a patient who is nearsighted. 
 In other words, we expect that ML predictions come with guarantees of \textit{integrity} (i.e., correctness)
on inputs whose integrity was itself not vouched for. Because these
inputs should not be classified by the model, it is unsurprising in hindsight
that it does not classify them accurately. 
This holds even if the model was perfectly learned, i.e. it recovered
the underlying data distribution.

As ML is increasingly applied to domains involving security or privacy considerations (e.g., intrusion detection\footnote{For instance, one could create a learning algorithm that models network behavior and predicts whether a particular set of packets contain the payload of an attack or not.}, high-frequency trading~\cite{kearns2013machine}, healthcare~\cite{esteva2017dermatologist}, ...), the assumption that training and test distributions are identical does not hold anymore: adversaries will intentionally find inputs such as the example described in the previous paragraph. In practice, adversaries may attempt to find attack variants that evade intrustion detecion, manipulate the inputs of high-frequency trading to cause them to issue disadvantageous transaction orders, or inspect healthcare models to recover sensitive data. As economist Goodhart once stated, “when a measure becomes a target, it ceases to be a good measure”. Here, we could derive that “when a model becomes a target, it starts to be vulnerable.”
This means that the model itself could be  the subject of an attack; or that it may be (a) presented with inputs that were manipulated during training and (b) asked to make predictions on inputs that it was not trained to predict on. 
Examples of such attacks have already been explored in the literature: e.g., model inversion~\cite{fredrikson2015model}, data poisoning~\cite{biggio2012poisoning} or model evasion~\cite{biggio2013evasion,szegedy2013intriguing}.

Importantly, potential misuses of learning-based systems are not limited to the ML model itself but also the computer system hosting this model. In other words, the security of a ML model can (1) be impacted by but (2) can also itself impact the security of the system that is deploying this very ML model. 

First, a ML model can only be as secure as the system that hosts it. This boils down to establishing a trusted computing base. For instance, the confidentiality of data analyzed by a ML model during training or inference could be jeopardized by a side-channel on the accelerator (e.g, FPGA, GPU, TPU, etc...) that runs the model.
Without having the whole system in mind, it could also be difficult to trace back the impact of training and test data on the model's behavior. This consideration, in many ways comparable to data provenance~\cite{buneman2000data}, is important when thinking about accountability in the context of ML---given that a number of attacks (e.g., poisoning, watermarking) attempt to modify the model's behavior without the defender's knowledge. We will come back to this issue later in our discussion of complete mediation and admission control. 
These two points suggest that in cases when establishing what should be included in the trusted computing base is difficult, a defense-in-depth approach to security from a system-wide perspective will likely also be relevant for learning-based systems. 

Second, the security of a system deploying ML can also depend on the security of the ML model itself. For instance, some  security properties such as availability only make sense in the context of the entire system but may depend on security properties of the ML component itself (e.g., integrity). Take the example of a software defined network (SDN) controller that integrates a ML model for intrusion detection. If the integrity of that intrusion detection model cannot be guaranteed, the availability of the SDN controller and of the entire network it manages may be affected as well.

\section{Revisiting Saltzer and Schroeder's principles}

Having established the importance of security and privacy for machine learning, we now compare ML security with traditional computer security. Specifically, we analyze the security and privacy of machine learning systems through the lens of Saltzer and Schroeder's principles~\cite{saltzer1975protection}. These principles, written largely in the context of cryptography research, not only ensure that our comparison with traditional computer security is systematic, but they also provide insights on what ML security is missing to achieve the level of provable security guarantees that cryptography does today. These 10 principles strive for simplicity and restriction. Indeed, these two characteristics generally make a security design less prone to error and easier to manage. This analysis leads us to identify three directions for future research that we explore in the rest of the presentation.

% , this will probably benefit from a system-wide perspective where authentication and authorization enforce all accesses to the model. 

\paragraph{Fail-safe defaults.} \textit{Base access decisions on permission rather than exclusion.} A precise characterization of deviations from the intended learning and inference behaviors should help implement fail-safe defaults that prevent the model from being exploited beyond its original purpose. 

At test-time, a conceptually simple example would have the system not reveal model predictions when they are made with low confidence. This is not a perfect mechanism in the sense that the adversary will likely be able to adapt its attack strategy to the specific threshold used to specify the minimum confidence needed, but it does increase the adversary’s cost. At training time, when learning is performed in an online fashion, rather than applying the updates systematically after each point is received, one could build an analog of \textit{sandboxes} for ML. For each batch received, the defender computes an updated candidate model and compares its performance on a holdout set of data to the current model. If the change in performance is not too important, the candidate model replaces the existing model. Otherwise, the candidate model and the corresponding batch of data are discarded by default. This makes the adversary's cost higher by forcing them to introduce changes gradually. 

\paragraph{Open design.} \textit{The design of security mechanisms should not be secret.} Existing ML research has already shown that obfuscation strategies that attempt to conceal information from attackers only provide limited security benefits and can be circumvented by motivated adversaries. 

One could for instance imagine that crafting adversarial examples in a black-box threat model (where the adversary does not have direct access to the model) is more difficult than mounting the same attack in white-box settings. However, that is rarely the case (i.e., the gains of having a secret design are quite limited for security) because of two reasons. First, a model that is originally a black-box may become a white-box later in the future. That is, an insider could leak the model or an adversary with access to a device deploying the model could reverse-engineer the device’ software to recover the model. Second, adversaries with access to a black-box model’s predictions are able to extract a substitute model (i.e., a local copy of the black-box model that is now accessible to the adversary) that can be used to mount other attacks (e.g., find adversarial examples). This extraction typically consists in sending carefully selected input queries to the black-box model and observing its predictions. Using the substitute model to find inputs that manipulate the black-box model then succeeds because we have observed that different ML models often share a large subspace of their error space: thus, attacks that the adversary finds on the extracted substitute model are also likely to be effective on the black-box model. 

\begin{figure}[t]
\centering
\includegraphics[width=0.75\textwidth]{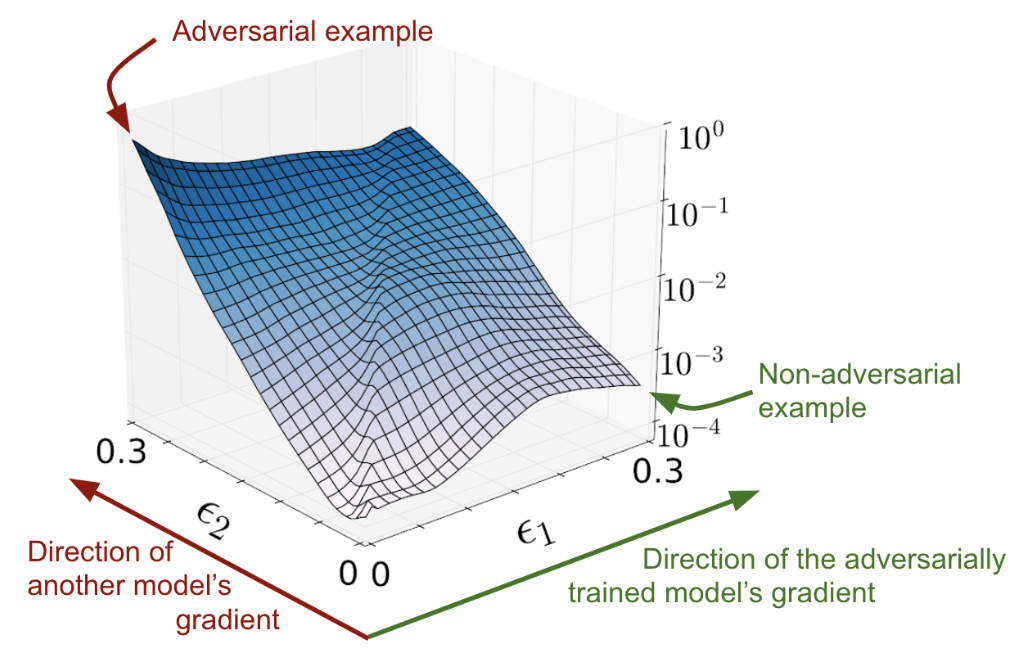}
\caption{Example of an attack that circumvents gradient masking. Learning models with adversarial training (i.e., injecting adversarial examples during training) often results in gradient masking: the model's gradients are no longer informative to find adversarial examples with the attack that was used to train the model. Yet, an adversary can adapt by considering the model whose gradients were masked as a black-box, and transfer adversarial examples found on a different model whose gradients are not masked. This illustration visualizes the loss of an adversarially trained model (vertical axis) on adversarial examples computed by evaluating the attack directly on the adversarially trained model or a different model. The origin corresponds to the unmodified test input $x$ and all other points ($\alpha, \beta)$ to $x+\alpha\varepsilon_1 + \beta\varepsilon_2$ where $\varepsilon_1$ is the direction computed on the adversarially trained model (right horizontal axis) and $\varepsilon_2$ on the other model (left horizontal axis). The attack is more successful (larger loss) when transferring adversarial examples found on the other model than it is when computing the attack on the adversarially trained model directly. See~\cite{tramer2017ensemble} for more details. }
\label{fig:gradient-masking}
\end{figure}

Another related example is the one of gradient masking. 
In particular, in the context of test-time attacks (e.g., adversarial examples), gradient masking~\cite{papernot2017practical} is a class of defense techniques that manipulate a model's gradients, which are often essential to an adversary's reconnaissance of the model while conducting a test-time attack. While this closed-design strategy increases the adversary's cost, an adaptive adversary can circumvent such defenses~\cite{papernot2017practical,athalye2018obfuscated} (see Figure~\ref{fig:gradient-masking}).

On the contrary, privacy mechanisms are for the most part designed to provide worst-case guarantees that hold regardless of the knowledge and capabilities of the adversary. This is the case of differentially private mechanisms. 

Contributing to high standards of open design, the ML security and privacy community consistently releases open-source code through libraries like CleverHans~\cite{papernot2016cleverhans} and FoolBox~\cite{rauber2017foolbox} or repositories for individual papers.

\paragraph{Separation of privilege.} \textit{Where feasible, a protection mechanism that requires two keys to unlock is more robust and flexible than one that allows access to the presenter of only a single key.} While somewhat specific to the cryptography research that motivated Saltzer and Schroeder's principles, separation of privilege is still applicable to ML and in particular distributed settings for ML. 
One prominent framework implementing this separation of privilege is federated learning where rather than collecting data centrally, a ML model is built by having clients compute model updates individually on their own data and then aggregating these local updates only~\cite{mcmahan2016communication}. Another example is the ESA architecture, where the encoding, shuffling and analysis steps provide privacy in the data pipeline~\cite{bittau2017prochlo}. One can possibly involve multiple parties in order to separate privileges (e.g., by having more than one entity responsible for shuffling the data before it is analyzed by a third party).  In a completely different threat model, one could also envision using an ensemble of models trained on independent data pipelines to reduce one's exposure to poisoning attacks.

\paragraph{Work factor.} \textit{Compare the cost of circumventing the mechanism with the resources of a potential attacker. }
For instance, designing CAPTCHAs that are difficult to solve for ML models makes little sense given the low cost of human labor needed to solve large numbers of CAPTCHAs~\cite{motoyama2010re}. 
While early ML security work on evasion attacks considered the cost expended by adversaries when mounting attacks (e.g., in adversarial learning~\cite{lowd2005adversarial}), this consideration has been absent of most recent efforts. This is most likely due to the surge of interest in image classifiers, for which inputs are often quasi-instantly and arbitrarily manipulated. However, domains like intrusion detection may require higher per-attack costs if the adversary needs to adapt to evolving feature engineering performed by the defender. 

Generally speaking, the adversary's work factor is also much lower when the defender commits to an approach first and the adversary makes its move second. For instance, it may be easier for defenders to remove backdoors inserted in ML models than it is to reliably watermark a model before it is released. Watermarking is an instance of poisoning where the defender intentionally poisons the model to have it respond to a specific trigger of their choice. Potentially, this could help the defender identify stolen copies of their model only by querying them. While similar techniques allow an adversary to insert a backdoor in a model, in the backdoor setting, the defender can manipulate the model once the attacker has committed to it. Instead, in the watermarking setting, the defender must commit first. While both settings may eventually result in an arms race, the defender is advantaged in the case of backdoors whereas the attacker is advantaged in the case of watermarking because it is currently often easier to remove correlations that result from poisoning than it is to insert them robustly (so long as the poisoned behavior is inserted outside of the training manifold).

% On the other hand, privacy-preserving ML does not make any assumptions about adversarial capabilities and thus matches this principle.

\paragraph{Psychological acceptability.} \textit{It is essential that the human interface be designed for ease of use, so that users routinely and automatically apply the protection mechanisms correctly.} This is particularly relevant to ML security given that ML models (and in particular deep neural networks) are commonly depicted as being black-boxes.\footnote{Whether that should be the case is the subject of another discussion.} Here, the psychological acceptability of security and privacy mechanisms will likely benefit from and be beneficial to other areas of research directions like interpretability~\cite{lipton2016mythos}. They share the goal of providing an explanation---that is easily accepted by a human---for why a model accepts or refuses to classify an input (e.g., because the model's confidence is high or low and the prediction likely to be respectively correct or incorrect). 

For instance, the Deep k-Nearest Neighbors approach~\cite{papernot2018deep} searches the training data for points whose internal representations are closest to the internal representations of the test point being classified. These training points not only contribute to explaining why the model is making a particular prediction, but their label can also serve as a way to reliably estimate the uncertainty of the prediction---which is useful to build a mechanism that rejects inputs manipulated by adversaries, as described in the fail-safe paragraph previously. 

Privacy mechanisms have long suffered from their complexity, which impacted their psychological acceptability and deployment. However, recent proposals like PATE~\cite{papernot2016semi} provide rigorous---yet presumably intuitive---guarantees. The approach partitions a sensitive training dataset in order to independently train an ensemble of ML models on each data partition. At test-time, the models are asked to predict by collectively voting for one of the possible labels. See Figure~\ref{fig:pate} for a diagram of the approach. Thus, if a PATE ensemble correctly classifies an input, it is because there is consensus among  ML models trained independently on different data partitions. This not only makes it easy to understand where the privacy guarantee comes  from: any given training point has a limited impact on the prediction made because it was used to train a single teacher.
Beyond the benefits to security, being able to explain a model's prediction also helps deploy ML in high-stake applications (e.g., healthcare) and can provide insights about the data: low performance of a private ML model on certain inputs may reveal the need to consider fairness. 

\paragraph{Economy of mechanism.} \textit{One should keep the design of security mechanisms as simple and small as possible.} Unfortunately, this is often difficult because many security problems in ML relate to the methods that underlie ML: because training objectives and algorithms rely on the assumption that training and test distributions are identical, it is more difficult to design robust ML models without making significant changes to them. Nevertheless, the principle still applies to ML security: for instance, when attention is given to details, adversarial training  can outperform many elaborate schemes for thwarting norm-constrained adversarial examples. Beyond the reduced risk associated with them, simple and concise security mechanisms are also more likely to enable open design and psychological acceptability (see above). In particular, it is important to keep the interfaces of security modules simple to ensure that they are used properly in conjunction with the rest of the system. For instance, when designing the attack API for CleverHans, we found that it is important to provide a simple and clear abstraction for ML model predictions: not doing so otherwise often led to incorrect implementations (e.g., with numerical instabilities). 

\paragraph{Complete mediation.} \textit{Every access to every object must be checked for authority.} This is relevant to ML security in several ways. First from a confidentiality and privacy standpoint, it is key to enforce access control to the model and its predictions. The model itself may constitute intellectual property~\cite{tramer2016stealing} or the basis for more elaborate attacks that recover sensitive information analyzed by the model during training~\cite{shokri2017membership,carlini2018secret}. Second, recall the data provenance question raised earlier. Complete mediation is also relevant to it: if the defender is unable to verify the integrity of its training or test data, it potentially exposes the model to poisoning, evasion or privacy attacks. It could also make it difficult to implement the failsafe defaults mentioned earlier. In both cases, the defender's task is to cope with data that is not from the distribution of interest, i.e. distribution drifts. We  come back to this later in Sections~\ref{sec:assurance} and~\ref{sec:auditing}.

\paragraph{Least privilege.} \textit{Every program and every user of the system should operate using the least set of privileges necessary to complete the job.} Beyond the system-wide considerations---which fall under the realm of traditional computer security, the ML paradigm offers new ways to implement the least privilege principle. For example, the entity serving a ML model may not be trusted with access to the data which the model is inferring on. In this trust model, one could leverage homomorphic encryption to guarantee that the test data and predictions made by the model can be accessed only by the user who submitted this test data for inference~\cite{graepel2012ml}. One could also imagine that a ML system would lose access to its training dataset once it has been trained.

\paragraph{Least common mechanism.} \textit{Minimize the amount of mechanisms common to more than one user and depended on by all users.} As mentioned previously, training data for an application can often be collected from several sources with varying degrees of integrity guarantees. For instance, a system may deploy mechanisms specific to detect poisoning attacks on data collected from the Internet or labeled by external operators, but data that was collected and labelled by a trusted party may not need to go through this potentially expensive verification procedure. Similarly, different users may call for varying degrees of privacy guarantees~\cite{abadi2017protection}: for example, a subset of patients may have opted-in to contribute their data for research purposes in a medical application. Homomorphic and non-homomorphic variants of the same system could also be used for data with varying confidentiality risks. Finally, if the defender is able to build robust ML models that are expensive to run inference for, it may choose to deploy these models to respond to queries which are untrusted and could be adversarial examples, while queries from trusted users could be answered by a less computationally intensive (or simply more accurate) but less robust model.

\paragraph{Compromise recording.} \textit{It is sometimes suggested that mechanisms that reliably record that a compromise of information has occurred can be used in place of elaborate mechanisms that completely prevent loss. }
Compromise recording is particularly difficult for ML systems because the relationship between test performance and training performance is poorly understood and limited formal tools are available to analyze it. 
For instance, being able to identify training points that resulted in a particular model prediction can serve as the basis for evaluating how robust a model is to data poisoning (e.g., with influence functions~\cite{koh2017understanding}) or designing mechanisms that reject test inputs which are not part of the expected training distribution (e.g., because they have limited support in the training data~\cite{papernot2018deep}). In another example, it is currently unknown whether it is possible to detect model extraction
under certain constraints placed on the adversary. This would reduce the need for complicated watermarking defenses, that are likely to come at the expense of performance for the model's expected users. We come back to this in our discussion of ML auditing.

\paragraph{} Three directions for future work emerge from this discussion of past efforts: model assurance and admission control, auditing ML and the need for formal security and privacy goals and frameworks that are better aligned with original ML goals. We discuss each of these in the rest of the presentation.

\section{Model assurance and admission control}
\label{sec:assurance}

Machine learning research has historically focused on providing and measuring expected performance of the system. A typical proxy for generalization error is to measure the accuracy of the model on test data. Instead, in security, we would like to provide worst-case guarantees: among a set of attack strategies, adversaries will always choose the one that benefits them the most. Practically speaking, this requires that we shift from testing to verification. For instance, we could measure how little a model memorizes specific training points (e.g., through metrics like exposure~\cite{carlini2018secret}) but this does not guarantee that the model does not memorize any training point.  

Unfortunately, formal verification techniques designed for programs do not directly apply to ML systems and thus need to be adapted. 
% This enables one to debug ML models as they are being trained. Once models are deployed, input verification  ensures that models only infer on inputs that are from the test distribution they were designed to handle. As a by-product, the system can revert to fail-safe defaults (see above) if input verification mechanisms achieve complete mediation.
At training time, model assurance would eventually enable one to establish with confidence that the ML system matches security requirements. At test time, admission control would consist in deciding whether input-output pairs obtained from a sandboxed execution of the model are admissible in the pool of answers. Both model assurance and admission control require however that a 
security (in the broad sense here, that also includes privacy) policy be specified.

\paragraph{Specifying privacy policies.} Finding the right language for specifying security and privacy policies in ML systems is a prerequisite for model assurance and admission control, yet it remains an open problem. 

As far as security (e.g., integrity) is concerned, we have an informal way to specify the desired requirements. For instance, we’d like the system to accurately model exactly the task	 which it was designed to solve. This implies that the system should be correctly implemented (e.g., not present any numerical instabilities), solve the end task (e.g., make correct predictions on all valid inputs) and only solve this end task (e.g, not present any backdoors or corrupted logic resulting from data poisoning). However, it is unknown how to formalize this intuitive set of requirements with precise semantics while avoiding ambiguity. 

In privacy, we have a much clearer picture where one can state both intuitively and rigorously what requirements a privacy-preserving ML system should match. Intuitively, the system’s behavior should not reflect any personal information included in the training dataset. Formally, one can use the framework of differential privacy to specify this requirement~\cite{dwork2006calibrating}. 

We will come back to this discussion at the end of the talk but for now we will assume that the security (and privacy) policy has been specified.

\subsection{Model assurance at training time}

Formal verification is a popular way of establishing security assurance, i.e., confidence that the system matches the  security requirements stated.

\paragraph{Coverage in machine learning.} Once security requirements have been established, one obstacle that prevents the direct application of formal verification techniques designed for programs is that they often rely on a notion of coverage. Intuitively, \textit{coverage} measures how much of the program's logic was exercised by the verification tool. A simple yet common coverage metric is the number of lines executed by a verification program. This metric does not apply well to ML because an inference pass through most modern implementations of ML models will likely execute all of the program's lines. In other words, the number of lines executed does not reflect how much of the model's logic has been verified because a lot of the computations involved are defined as matrix computations. To circumvent this limitation, a first approach introduced neural coverage as the number of neurons whose activation is above a threshold~\cite{pei2017deepxplore}. Unfortunately, this does not always completely reflect the logic learned by a model, which is hardcoded by complex arrangements of multiple neurons. Another recent proposal instead  considers layer activations as a whole and measures the distance between different activation vectors to measure progress~\cite{odena2018tensorfuzz}.
Future work will have to find the right abstractions for representing components of a neural network's logic, which will likely also help build more interpretable and robust models. One could for instance ask how the dynamic routing of activity vectors would help with identifying such logic elements in capsules networks~\cite{sabour2017dynamic}. 

In addition to covering the entire logic implemented by the system, it is important to verify the entire set of valid inputs to the system. In ML this can be more difficult than expected because this set is often poorly defined and part of the motivation behind applying ML is to identify this manifold. For instance, it is possible to define the set of images using a range of pixel values but this is different from the manifold of natural images. This question is important because verification tools often need a dataset to bootstrap an heuristic. Finding datasets representative of the expected task distribution can be hard and is necessary to avoid some of the pitfalls of testing.

\subsection{Test-time admission control of input-output pairs}

In traditional access control, it is common to implement a reference monitor. The reference monitor mediates all requests from sources to guard a specific resource. In our case, the sources would be users, requests prediction queries and the resource would be the ML model itself or its prediction. Queries that we would like to reject include those that were crafted to have the model mispredict (e.g., adversarial examples) but also those that are part of reconnaissance efforts (e.g., model extraction) to mount other attacks. 

However, because authentication is often weakly enforced when queries are made to a ML model, as is the case for search engines, implementing a reference monitor is difficult and admission control is essential to prevent adversaries from exploiting vulnerabilities that were not addressed by model assurance. Indeed, just like a program can be policy-compliant yet still have buffer overflows, one could imagine that a ML model may still exhibit undesired behavior despite passing model assurance. This could be the case if the security policy does not apply to a zero-day attack. Hence, given an input and an output, we’d like to be able to know whether we admit the input-output pair into our pool of answers. This is difficult in ML because the underlying distribution is unknown.

% For instance, while there is no universal and effective way to prevent spammers from sending spam, spam is no longer a problem because a large portion of email is sent through companies like Microsoft or Google that have ways to authenticate users. If similar ways to authenticate source of requests cannot be implemented in ML applications and data provenance is difficult, we will have to resort to other techniques for verifying that inputs processed by the model are benign. \todo{might want to change to search engine}

\begin{figure}[t]
\centering
\includegraphics[width=0.9\textwidth]{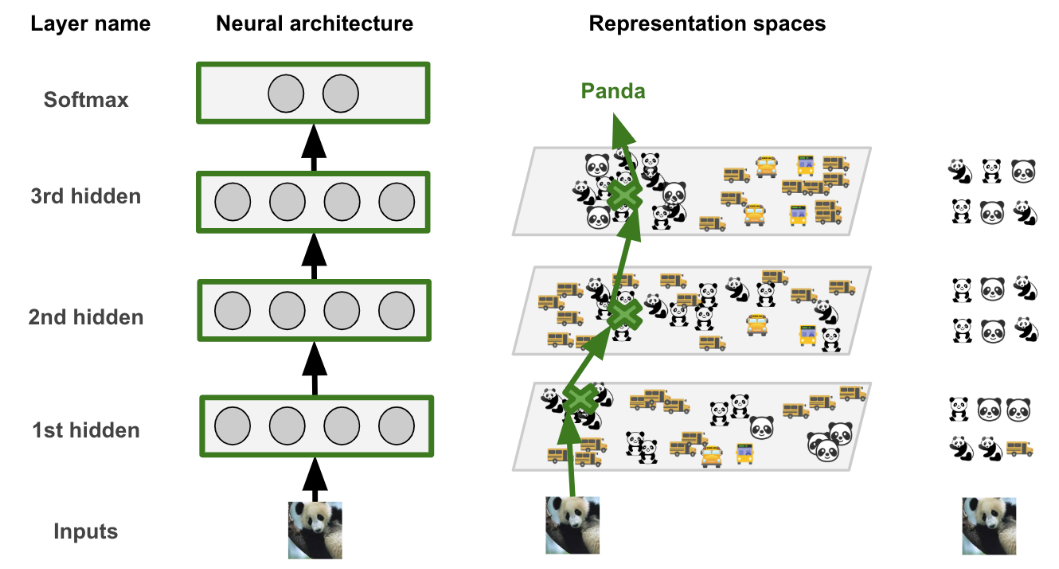}
\caption{The Deep k-Nearest Neighbors estimates support in the training data for a particular prediction through a nearest neighbors search over the internal representation spaces. This serves as the basis for admission control of input-output pairs based on the homogeneity of the support, which is a proxy for the model's uncertainty. See~\cite{papernot2018deep} for more details. }
\label{fig:dknn}
\end{figure}

In essence, we would like to verify that we only use the model on the intended test distribution. One approach is to run the model in a sandbox and analyze  the input-output pair before revealing it to the user. 
For instance, if we can define a well-calibrated estimate of the model's
uncertainty, we can admit input-output pairs only if the model's confidence is sufficiently high.
Evaluating these defense strategies necessarily involves a tradeoff between performance on legitimate and adversarial inputs, as increasing the true positive rate (adversarial examples marked as malicious) often undesirably increases the false positive rate (legitimate inputs marked as malicious)~\cite{anonymous2019evaluation}.
Confidence is a difficult metric to use because it can be manipulated by adversaries. For instance, in a modern neural network, an adversary can typically modify the input to modify the magnitude of the pre-softmax activations (i.e., logits) and thus arbitrarly control the model's confidence. This is because high confidence in a particular label only suggests that the model believes that there are no other possible labels for the given input based on the training data it has analyzed. However, this does not mean the training data is relevant to classify this particular input. 

Hence, estimating  support in the training data when making a prediction can serve as the basis for coming up with more reliable uncertainty estimates. This is for instance the approach taken in the Deep k-Nearest Neighbors, depicted in Figure~\ref{fig:dknn}, where support from the training data is evaluated by searching for training points whose internal representations are closest to the ones of the test input being classified~\cite{papernot2018deep}. The labels of these nearest training points is then analyzed: when they are homogeneously supporting the  predicted label, uncertainty in the prediction is low; instead, when there are several different labels in the nearest training data, uncertainty is high. Note that characterizing support from the training data also has the benefit of informing active learning procedures: when weak support is identified for a particular input, this can be prioritized in the labeling strategy. Other approaches that seek to provide reliable uncertainty estimates include Bayesian deep learning~\cite{kendall2017uncertainties}.

\section{Towards Auditing ML}
\label{sec:auditing}

As is the case for search engines or spam filtering---where authentication and authorization are weakly enforced, auditing will likely be an important component of ML security~\cite{lampson2004computer,john2010searching}.
The two components of auditing---better compromise recording and a posteriori analysis of these logs---would partly alleviate incomplete assurance through the identification of threats both proactively and reactively. Specifically, it may help with increasing the work factor of adversaries that wish to go undetected as well as improve the psychological acceptability of ML security mechanisms if a complete record of compromise is achieved. 
Combined with techniques like sandboxing, auditing may improve the security of ML systems in an  manner orthogonal to other security-related efforts.

The first step towards designing an audit system is to identify which information should be logged and analyzed. This decision should be made in light of the security requirements, which inform the process of identifying all of the information that is relevant to detect and analyze malicious behavior that violates the security policy. An analyzer capable of deducing violations of the security policy then parses the information contained in these logs.

Designing auditing systems for ML is still largely an open problem. Hence, in the following we give examples where auditing would improve how practical attacks against learning or inference are handled and what information could be collected in logs for that purpose.

\begin{figure}[t]
\centering
\includegraphics[width=\textwidth]{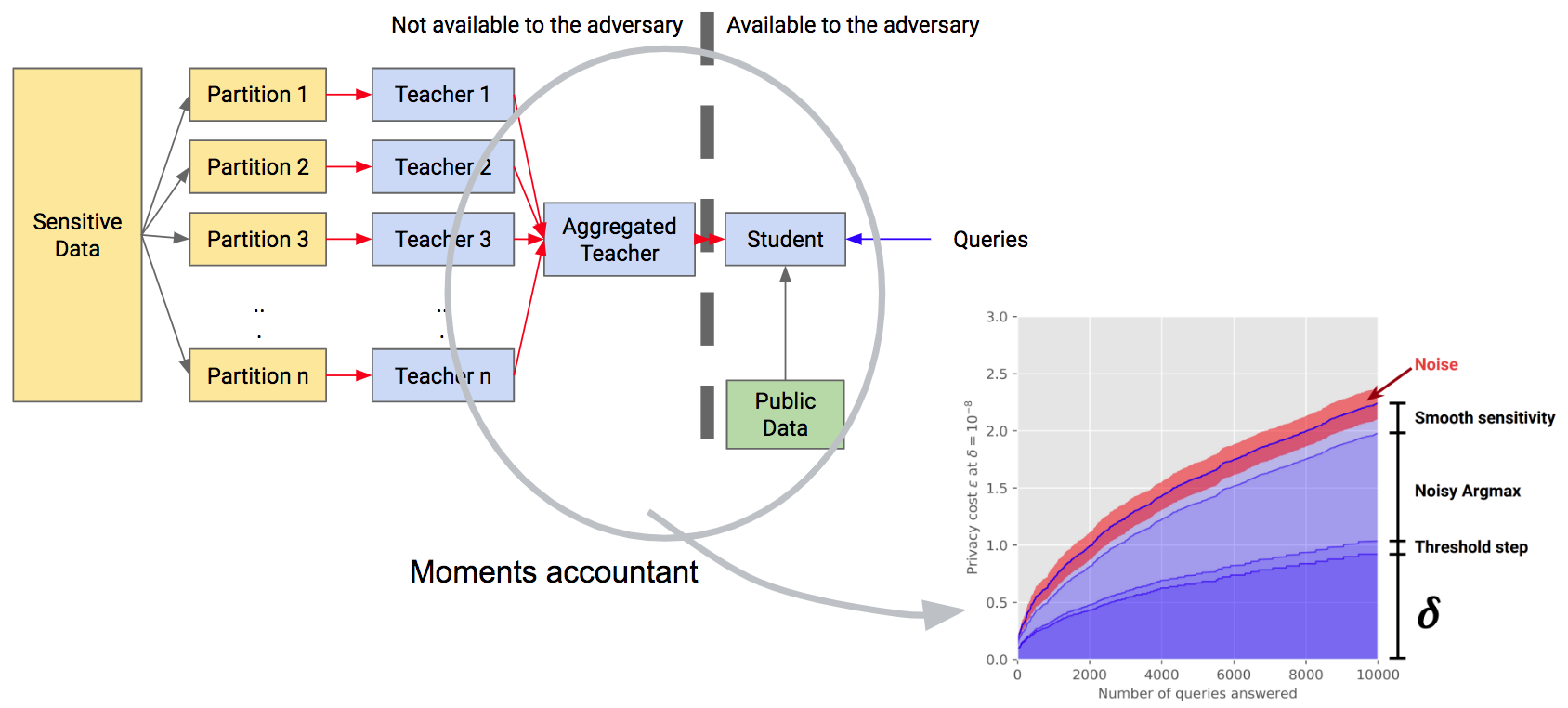}
\caption{Example auditing of a learning algorithm's privacy properties. The moments accountant not only analyzes the votes of teachers for each training label given to the student, the privacy analyzer also breaks down privacy expenditure by components of the overall approach. See~\cite{papernot2018scalable} for more details. }
\label{fig:pate}
\end{figure}

\paragraph{Auditing the learning algorithm.} Some forms of auditing systems exist in the privacy realm, most likely because the security requirements are rigorously formalized within the framework of differential privacy. For instance, the moments accountant~\cite{abadi2016deep} can be seen as an analyzer for logs produced by privacy-preserving algorithms like DPSGD or PATE. The moments accountant performs data dependent analysis to estimate whether the privacy budget was exceeded and reports these results to the user training the model. In the example of PATE, the histograms of teacher votes are saved for each aggregated label returned to the student. This logging is performed during training. Once the student is done training, the moments accountant can analyze these logs and give insights on how the privacy budget was spent during training (see Figure~\ref{fig:pate}). 

\paragraph{Auditing the inference algorithm.} Techniques that characterize the relationship between train and test behavior are likely to produce useful information to collect when auditing a learner during inference. 
In the context of privacy, logging user queries made for points that are exactly found in the training data (or very close to training data) could be the basis for an analyzer that identifies membership inference attacks if a large number of such queries is logged within a small time window. In some ways, watermarking could be the basis for collecting logs of model predictions made beyond the intended deployment systems and for instance detect model theft (assuming the defender has the ability to monitor the systems potentially hosting stolen copies of the model). More generally, making it more difficult for adversaries to perform the reconnaissance necessary to reverse engineer and extract models is a strong use case for auditing. However, it is unclear at this point how one could distinguish queries made during reconnaissance from legitimate user queries. 

\paragraph{} Obviously, logs collected to support auditing will be targeted by adversaries. It is important to insure they are tamperproof. This however falls under the realm of traditional integrity consideration and is not specific to ML.

\section{Formal frameworks that align ML goals with security and privacy goals}

Previously, we left the discussion of formal specifications of security and privacy policies on the side. However, precise definitions of policies are a prerequisite for model assurance, admission control and auditing.

\paragraph{A comparison with cryptography.}Cryptography made a lot of progress once the security game (which includes the adversary’s capabilities and goals) was identified and defined formally. For instance, if we look at protocols and how they evolved from SSL, to TLS 1.0, 1.2 and recently 1.3, it is clear that the increasing use of scientific thinking, provable guarantees, formal analysis and hard mathematical reasoning was beneficial to the security of the TLS 1.3 protocol. One differentiating  characteristic of cryptographic systems is that they are---for the most part---expressed mathematically. This facilitates the formal analysis of their security properties. However, a large part of ML architectures and algorithms can also be expressed mathematically. Hence, we ask here: \textit{are ML systems more amenable to formal specification of security and privacy goals than traditional computer systems because a large part of the system can be expressed mathematically?} One caveat that may prevent us from answering this question positively lies in the lack of precise generalization bounds between training and test-time performance.

\paragraph{Differential privacy.} A great example of a formal framework that supports both intuitive and rigorous specifications of a policy is differential privacy~\cite{dwork2006calibrating}. Intuitively, an algorithm is differentially private if its behavior is indistinguishable to an adversary regardless of the fact that any of the individual data points were included or not. This can be expressed formally. Indeed an algorithm $M$ (here, this would be a learning algorithm) is said to be differentially private if for any neighboring datasets $d$ and $d'$ (that is, they have a Hamming distance of 1) and any possible output $S$, the following holds: 
$$Pr[M(d)\in S] \leq e^\varepsilon Pr[M(d') \in S]$$
Here, the strength of the privacy protection is measured by $\varepsilon$: smaller values of $\varepsilon$ correspond to stronger privacy guarantees.

In addition to not making any assumptions about the adversary (which means it holds in the face of new attacks that may be discovered in the future), differential privacy is great because it aligns worst-case privacy guarantees with average-case generalization. 
For instance, in follow-up work on PATE, we found that we could make a simple modification to how individual teacher votes are aggregated, which results in simultaneous improvements to both the accuracy and the privacy  of the approach. Specifically, we explicitly required the consensus among teachers to be very high before revealing the corresponding aggregated label to the student (note how this recalls the fail-safe defaults principle discussed earlier). First, we found that this improves the mean accuracy of labels presented to the student, and as a consequence the accuracy of the student itself: if the consensus among teachers in the ensemble is high, the label is more likely to be correct. Second, this also reduced the overall privacy budget spent to create these labels because each label is very cheap in terms of privacy: it is very unlikely for a single teacher to change the outcome of aggregation. 

\paragraph{Towards a similar framework for security.} If we'd like to define a similar framework for security, we'll have to answer the following questions: 

\begin{itemize}
    \item \textit{Should guarantees be formulated with respect to the training data, algorithm or both?} Differential privacy is a property of the algorithm and not of the data. Yet, it may make sense to include the data in a security policy: for instance, it is likely harder to train a robust model on a dataset that is not linearly separable than on a dataset that is linearly separable. 
    \item \textit{Should the framework encompass training and test time adversaries?} One argument in favor of including both  in a unified framework is that it would allow us to consider dynamics between training-time attacks and test-time attacks: for instance, how does defending against adversarial examples impact robustness to poisoning attacks? These dynamics have been fairly unexplored by our community so far.
    \item \textit{How can we provide domain-agnostic formalism?} There has been a lot of attention to vision classifiers in our field over the past few years. This led us to define adversaries using formulations that only make sense in the image domain (e.g., $\ell_p$-norm based constraints on input perturbations). Considering other domains both in our work and in efforts to formulate security policies will thus likely lead to additional insights and more fundamental definitions of robustness in machine learning.
\end{itemize}

\section{Conclusions}

The main takeaways from this presentation are the following three points: 

\begin{itemize}
    \item \textbf{Efforts need to specify ML security and privacy policies.} What is the right abstraction and language to formalize security and privacy requirements with precise semantics and no ambiguity? In particular, we did not discuss how writing security policies for ML systems is different (or not?) from writing security policies for classic computer programs and systems. There is usually a difficult tradeoff between writing policies in sufficiently abstract languages for humans to understand them but at the same time in a sufficiently low abstraction that machines can also understand. This is particularly important if we don't understand how the predictions or learning happens.
    \item \textbf{Admission control and auditing may address lack of assurance.} Future work should consider how sandboxing, input-output admission control and compromise recording help secure ML systems when data provenance and assurance is hard.
    \item \textbf{Security and privacy should strive to align with ML goals.} While formulating the security and privacy requirements, we should take particular care in thinking how they relate to important ML goals. For instance: how do private learning and robust learning relate to generalization? how does poisoning relate to learning from noisy data or distribution drifts?
\end{itemize}

\section*{Acknowledegments}

The author would like to thank the AISec organizers for inviting him to present
this keynote: Sadia Afroz, Battista Biggio, Yuval Elovici, David Freeman and Asaf Shabtai. The basis for the analysis presented, i.e., following principles distilled by Saltzer and Schroeder, is inspired by a previous analysis of DPSGD and PATE, originated by Mart\'in Abadi in an invited paper at CSF 2017~\cite{abadi2017protection}. The format for this technical report was inspired by the NIPS 2016 tutorial on GANs by Ian Goodfellow~\cite{goodfellow2016nips}. The author would like to thank Mart\'in Abadi, Nicholas Carlini, \'Ulfar Erlingsson, Nicholas Frosst, Ian Goodfellow, Ilya Mironov and Florian Tram\`er for fruitful discussions.

\end{document}